\documentclass[
 aip,
rsi,
amsmath,amssymb,
reprint,%
]{revtex4-1}

\usepackage[breaklinks=true,colorlinks=true,linkcolor=blue,urlcolor=blue,citecolor=blue]{hyperref}
\usepackage{graphicx}
\usepackage{epstopdf}
\usepackage{color}
\usepackage{graphics}
\expandafter\let\csname equation*\endcsname\relax
\expandafter\let\csname endequation*\endcsname\relax
\usepackage{amscd}
\usepackage{times}
\usepackage{float}
\usepackage{array}
\usepackage{blkarray}
\usepackage{booktabs}
\usepackage{lipsum}
\usepackage[mathscr]{euscript}
\usepackage{dcolumn}
\usepackage{bm}

\begin{document}
\preprint{AIP/123-QED}

\title[Complex Pathways to Cooperation Emergent from Asymmetry in Heterogeneous Populations]{Complex Pathways to Cooperation Emergent from Asymmetry in Heterogeneous Populations}

\author{Hao Guo}
\affiliation{Department of Computer Science and Technology, Tsinghua University, Beijing 100084, China}

\author{Chen Shen}
\affiliation{Faculty of Engineering Sciences, Kyushu University, Fukuoka 816-8580, Japan}

\author{Rongcheng Zou}
\affiliation{School of Automation, Northwestern Polytechnical University, Xi'an 710072, China}

\author{Pin Tao}
\affiliation{Department of Computer Science and Technology, Tsinghua University, Beijing 100084, China}


\author{Yuanchun Shi}
\affiliation{Department of Computer Science and Technology, Tsinghua University, Beijing 100084, China}

\author{Zhen Wang}
\thanks{Corresponding author: zhenwang0@gmail.com}
\affiliation{School of Artificial Intelligence, Optics and Electronics (iOPEN), Northwestern Polytechnical University, Xi'an {\rm 710072}, China}

\author{Junliang Xing}
\thanks{Corresponding author: jlxing@tsinghua.edu.cn}
\affiliation{Department of Computer Science and Technology, Tsinghua University, Beijing 100084, China}

\begin{abstract}

Cooperation within asymmetric populations has garnered significant attention in evolutionary games. This paper explores cooperation evolution in populations with weak and strong players, using a game model where players choose between cooperation and defection. Asymmetry stems from different benefits for strong and weak cooperators, with their benefit ratio indicating the degree of asymmetry. Varied rankings of parameters including the asymmetry degree, cooperation costs, and benefits brought by weak players give rise to scenarios including the prisoner's dilemma ($PDG$) for both player types, the snowdrift game ($SDG$), and mixed $PDG-SDG$ interactions. Our results indicate that in an infinite well-mixed population, defection remains the dominant strategy when strong players engage in the prisoner's dilemma game. However, if strong players play snowdrift games, global cooperation increases with the proportion of strong players. In this scenario, strong cooperators can prevail over strong defectors when the proportion of strong players is low, but the prevalence of cooperation among strong players decreases as their proportion increases. In contrast, within a square lattice, the optimum global cooperation emerges at intermediate proportions of strong players with moderate degrees of asymmetry. Additionally, weak players protect cooperative clusters from exploitation by strong defectors. This study highlights the complex dynamics of cooperation in asymmetric interactions, contributing to the theory of cooperation in asymmetric games.

\end{abstract}


\maketitle

\begin{quotation}
Cooperation is a fundamental aspect of human society, and extensive research has been devoted to understanding its emergence. However, exploration of evolutionary games in asymmetric interaction scenarios has been relatively limited. This study aims to investigate the evolution of cooperation in populations comprising both weak and strong players, characterized by asymmetric benefits. Depending on the rankings of parameters associated with the degree of asymmetry, cooperation cost, and the benefit to weak players, both types of players can engage in the prisoner's dilemma game, the snowdrift game, or a mixed situation where one type of player participates in the prisoner's dilemma game while another plays the snowdrift game. Our findings underscore the indispensable role of the interplay between population composition and the degree of asymmetry. It reveals that in asymmetric interactions, defection is dominant in well-mixed populations when strong players are involved in prisoner's dilemma games. However, cooperation increases if strong players engage in snowdrift games. On a square lattice, optimal global cooperation levels occur at moderate asymmetry levels. This enhanced cooperation on a square lattice can be attributed to the significant role weak players play in sustaining cooperative clusters.
\end{quotation}

\section{Introduction}

Cooperative behavior is a widespread phenomenon observed in both human and non-human societies \cite{2001Collective, apicella2012social}, manifesting in various interactions, such as addressing climate change challenges \cite{andrews2018high, wang2020communicating} and mitigating epidemic outbreaks \cite{helbing2015saving, wang2016statistical}. Despite its ubiquity, a fundamental question remains: why do individuals incur costs to help unrelated individuals \cite{schmid2021unified, roberts2008evolution}? Game theory, whether applied in evolutionary or social sciences contexts, has played a pivotal role in uncovering fundamental mechanisms of cooperation, as well as exploring various adaptations and nuances that arise from them \cite{dong2019competitive, guo2020novel, mathew2009does, wang2018replicator}. Typical examples include direct and indirect reciprocity, network reciprocity, group selection, and kin selection \cite{nowak2006five}.

To understand the persistence of cooperation, evolutionary game theory has generally extended from basic symmetric interaction to asymmetric scenarios \cite{smith1974theory, gaunersdorfer1991dynamics}. Among the crucial symmetric subclasses in evolutionary game theory is the two-player two-strategy game, where players simultaneously choose between being a cooperator or a defector \cite{szabo2007evolutionary}. In this setup, a cooperator receives a reward, denoted as $R$, if it encounters another cooperator but gets a sucker's payoff, denoted as $S$, when it encounters a defector. On the other hand, a defector receives temptation, represented by $T$, when it faces a cooperator, and punishment, denoted as $P$, when it encounters another defector. The high-profile games in biological are the prisoner's dilemma game ($PDG$), where the ranking satisfies $T>R>P>S$, and the snowdrift game ($SDG$), where the ranking satisfies $T>R>S>P$ \cite{doebeli2005models}. These two symmetric games also fall under the category of social dilemmas \cite{ dawes1980social, macy2002learning}, wherein mutual cooperation yields the highest collective payoff compared to other configurations. Building upon the foundation of these two symmetric games, a notably intriguing direction is to broaden the investigation to include asymmetric scenarios \cite{wang2014different, mcavoy2015asymmetric}. Furthermore, the exploration of multigame environments and the impacts arising from unequal player dynamics are garnering considerable interest \cite{szolnoki2014coevolutionary, szolnoki2016leaders, perc2008social}. In this context, our objective is to extend the fundamental models of the $PDG$ and $SDG$, examining the evolution of cooperation in scenarios where inequality is factored into both structured and unstructured populations.

Asymmetry is actually ubiquitous. Even among the members of the same species, there are variations in sex, strength, intellect, wealth, environment, and history of interactions that may cause differences in ability and behavior~\cite{wolf2012animal, dell2012evolutionary, szolnoki2016leaders, guo2023third}, with considerable consequences for ecology and evolution. Examples include human hunt-gather game \cite{smith2010wealth, apicella2012social} and power (horizontal) inequality \cite{boix2010origins, canelas2018horizontal}. Asymmetry can arise from inherent diversity among individuals or power struggles that lead to further disparities in various aspects of life \cite{perc2008social, 2011Dictator}. The effect of asymmetry on cooperation has been explored through two classical assumptions: ecological and genotypic setups \cite{mcavoy2015asymmetric}. Ecological asymmetry arises from the differences in the location or environment of the participants, while genotypic asymmetry stems from variations in the characteristics or attributes of the individuals themselves. Furthermore, the ability of human behavior to shape the environment has been recognized in evolutionary games with environmental feedback, revealing the impact of such feedback and the rate of ecological changes on the strength of social dilemmas \cite{hauert2019asymmetric}. Additionally, the study of sub-population games offers an opportunity to explore cooperative behavior in the presence of asymmetry \cite{Szolnoki2017Evolutionary, guo2020dynamics}. The structure of social interactions, whether unidirectional or bidirectional, also plays a pivotal role in shaping the evolution of cooperation \cite{su2022evolution}.
Another model commonly used to study unequals in cooperation is the linear public goods game. In this game, incorporating individuals with slightly unequal endowments can potentially promote the evolution of cooperation, especially when the game is played repeatedly \cite{hauser2019social}. These findings highlight how asymmetries provide valuable insights into shaping cooperative behavior.

\begin{figure*}
	\centering
	\includegraphics[scale=0.7]{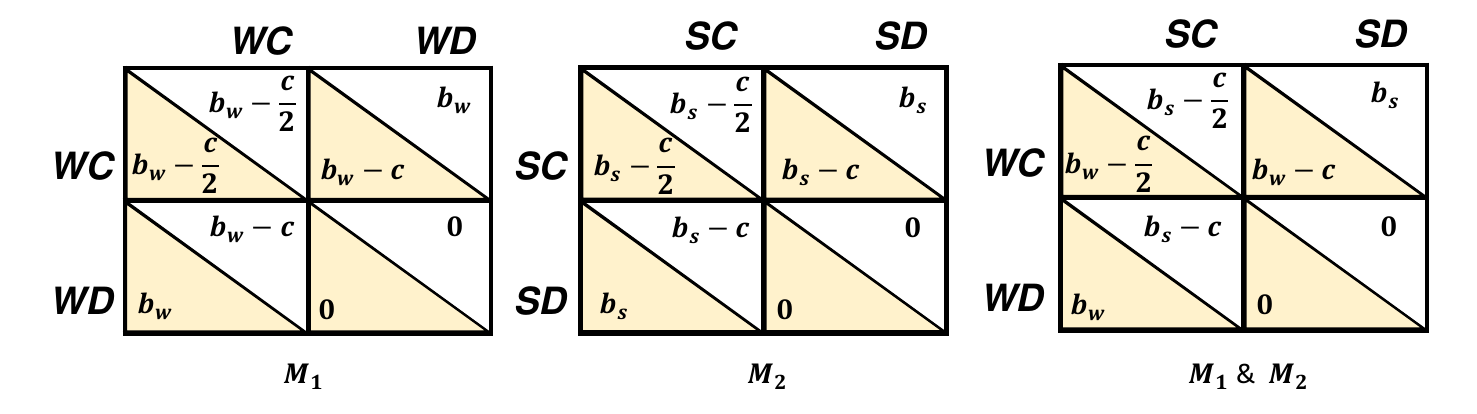}
	\caption{ {\bf Games in populations with inequality.} The matrices from left to right represent the game model played between two weak players, between two strong players, and between weak and strong players. Note that when two cooperators encounter, they evenly divide the associated cost. The resulting payoffs, occupying the yellow area of the first and second columns, can be effectively represented by $2 \times 2$ matrices denoted as $M_1$ and $M_2$, respectively. However, an inherent asymmetry arises when two players from distinct sub-populations interact.}
	\label{Sketch}
\end{figure*}

Motivated by these studies, our research expands upon the foundational models of $PDG$ and $SDG$ by incorporating populations with strong and weak tags. This allows for interactions between individuals from distinct sub-populations. In the game model, players have the choice between cooperation and defection. Cooperation yields mutual benefits, whereas defectors yield no benefits. If both players cooperate, they share the cooperation cost; otherwise, cooperators bear the cost alone. The asymmetry lies in the benefits of strong and weak cooperators. To quantify this disparity, we introduce the concept of asymmetric degree, differentiating strong players, who gain more from cooperation, from weak players, who gain relatively modest benefits. Considering the existing research indicating that human players often adjust their strategies based on social learning \cite{sigmund2010social}, our study incorporates a social learning rule in the strategy update phase. Specifically, players have the opportunity to adopt strategies not only from their own sub-population but also from different sub-populations. During this phase, individual population tags remain unchanged. Our research delves into the dynamics of cooperation under the interplay between asymmetry and population composition in structured and unstructured populations. For well-mixed populations, we find that global cooperation tends to rise with an increasing number of strong players participating in $SDG$. Notably, when strong players are involved in $SDG$ games, and their proportion is relatively low, strong cooperators often become predominant within the strong sub-population. Conversely, as the proportion of strong players in the population increases, the rate of strong cooperation among these players tends to decrease. On the other hand, in a square lattice, the optimal level of cooperation emerges at intermediate proportions of strong players with moderate asymmetric degrees, regardless of whether it is a $PDG-SDG$ or $SDG-SDG$ mixed scenario. Meanwhile, weak players guard cooperation clusters against exploitation by strong defectors. We therefore shed light on the intricate dynamics of cooperation in asymmetric interactions, highlighting the role of inequality and population composition in shaping human cooperation.

\begin{figure*}
	\centering
	\includegraphics[scale=0.6]{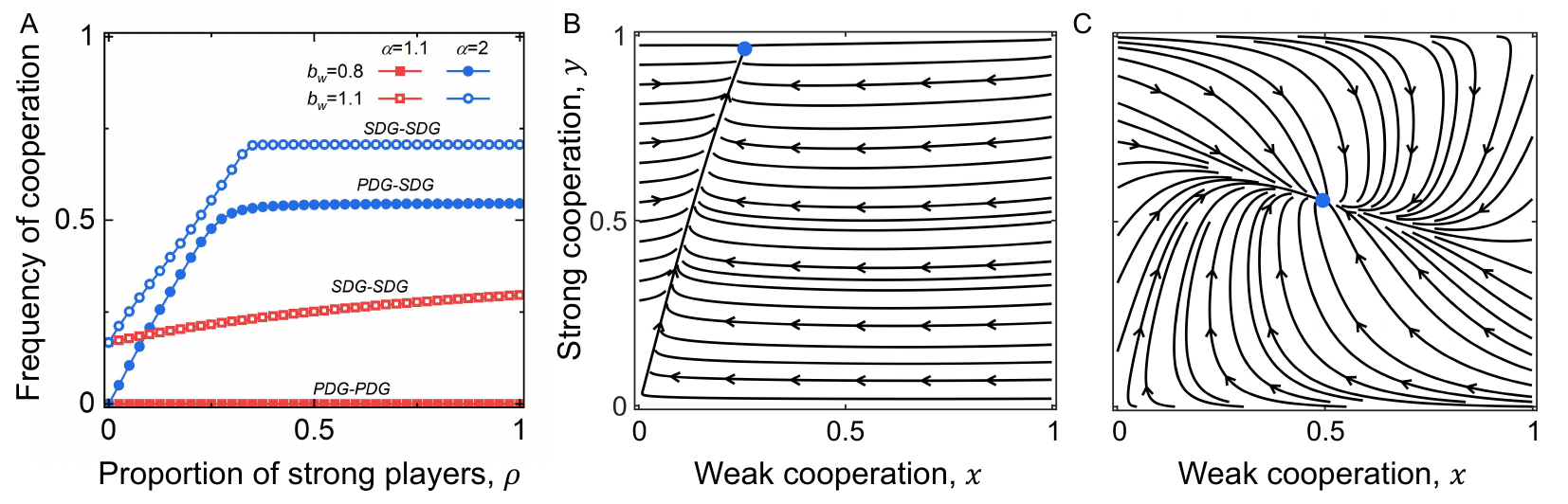}
	\caption{ {\bf Pure snowdrift game stimulates the highest cooperation in a well-mixed population.} A, shown are the steady-state levels of global cooperation as a function of the proportion of strong players, characterized by the $(b_w, \alpha)$ pair. The findings demonstrate that, in comparison to a mixed population, a population consisting solely of strong players facilitates a more pronounced evolution of cooperation. B, with a small proportion of strong players, strong cooperators exert dominance within the strong sub-population, thereby fostering the evolution of weak cooperation. The parameters are fixed as $b_w=0.8$, $\alpha=2$, and $\rho=0.2$. C, global cooperation increases as strong players increase. The blue solid dots represent attracting solutions. The parameters are fixed as $b_w=0.8$, $\alpha=2$, and $\rho=0.8$.}
	\label{wellmix}
\end{figure*}

\section{Materials and methods}

\subsection{Games in populations with inequality}

Our study incorporates tags into the basic $PDG$ and $SDG$, leading to an asymmetric game. In the game, each player has two options in the strategy set $\mathcal{S}=\{C, D\}$, where $C$ represents cooperation and $D$ represents defection. Cooperation provides a benefit $b$ to both the cooperator and the opposing individual, but also begets a cost $c$ if the opponent defects, and a cost $\frac{c}{2}$ if the opponent cooperates. In terms of the payoff matrix elements, this corresponds to $R=b-\frac{c}{2}$ for playing mutual cooperation, $T = b$ for defection against cooperation, $S = b - c$ for cooperation against defection, and $P = 0$ for mutual defection:

\begin{equation} 
\left( \begin{matrix} b-c/2 & b-c \\ b & 0
\label{matrix}
\end{matrix}
\right). \end{equation}

The parameter $b$ stands for the benefit and in our case helps to differentiate between two types of players and, consequently, games  (see the sketch in Fig.~\ref{Sketch}). Specifically, $b_w$ denotes benefit for weak players and $b_s$ denotes benefit for strong players. The parameter $b_s$ is a factor of $b_w$ and $\alpha$; $b_s = \alpha b_w$, where $\alpha$ is inequality degree. By setting $\alpha>1$, we ensure that strong players consistently receive a larger benefit $b_s$ compared to weak players' benefit $b_w$.
When cooperators encounter each other, they share the cost evenly and engage in either game $M_1$ or $M_2$ based on their tag (see Fig.~\ref{Sketch}). Here, the parameter $b$ determines whether player payoffs fall within the domain of the prisoner's dilemma or snowdrift game. Specifically, if $b \in (c/2,c)$, we have the $PDG$ for $T>R>P>S$, and if $b \in (c,+\infty)$, we encounter the $SDG$ for $T>R>S>P$. On the other hand, when interacting players come from different sub-populations, they participate in an asymmetric game. Specifically, the weak and strong players follow games $M_1$ and $M_2$, respectively (see the third column of Fig.~\ref{Sketch}). As $b_s \geq b_w$ holds constantly, there exist three distinct mixed game scenarios. (i) The $PDG-PDG$ scenario refers to both weak and strong players participate in $PDG$, implying that $\frac{c}{2}<b_w<b_s<c$. (ii) The $PDG-SDG$ scenario involves weak players participating in $PDG$ while strong players participate in the Snowdrift Game ($SDG$), indicating that $\frac{c}{2}<b_w<c<b_s$. (iii) The $SDG-SDG$ scenario comprises both weak and strong players participating in $SDG$, i.e., $c<b_w<b_s$.
In our study, without a specific declaration, the cost $c$ is fixed as $1$ throughout the paper.

\begin{figure*}
	\centering
	\includegraphics[scale=10]{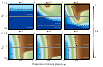}
	\caption{ {\bf Mixed population supports higher levels of cooperation in square lattices.} Showns are the steady-state levels of cooperation as a function of benefit for weak players $b_w$, asymmetry degree $\alpha$, and the proportion of strong players $\rho$. A, cooperation emerges in $SDG$ when both weak and strong players engage in the same symmetric game. B, cooperation emerges when even weak players participate in the $PDG$, provided strong players engage in $SDG$. Particularly, in the $SDG-SDG$ scenario, global cooperation attains higher levels in a mixed population compared to a pure weak or strong population. C, the optimal cooperation in mixed population occurs in $PDG-SDG$ scenario. D, optimal cooperation disappears in the $SDG-SDG$ scenario when $\alpha=3$. E, higher values of $\alpha$ appose payoff of strong players to the domain of harmony game, where the social dilemma is non-existent. Hence, peak cooperation levels are skewed towards higher densities of strong players. F, global cooperation increases monotonously with the proportion of strong players when the asymmetric degree is sufficiently large. The colorbar represents the frequency of global cooperation.}
	\label{Fig1}
\end{figure*}

\subsection{Population setup}

This study explores two main population structures: a well-mixed population and a square lattice with periodic boundaries. Individuals are represented as vertices, and edges represent their interactions. Specifically, if there is a link between individual $i$ and $j$, we denote it as $a_{ij} = 1$, and if there is no link, we set $a_{ij} = 0$. The players in the population are enumerated as $\mathcal{P} = \{1, 2, 3, \dots, N\}$, where $N$ represents the total number of individuals.
In a well-mixed population, every individual is connected to each other, i.e., $\forall a_{ij}=1, i \neq j$, allowing interactions between any two individuals.
We firstly consider an infinitely large well-mixed population, where $N \rightarrow +\infty$. 
To simplify notation, we use $\omega = 1 - \rho$ and $\rho$ to denote the proportions of weak and strong players in the population, respectively. Let $x$ represent the frequency of cooperation in the weak sub-population, which takes values in the range $[0, 1]$, and $y$ represents the frequency of cooperation in the strong sub-population, also ranging from 0 to 1. Consequently, the total frequency of cooperation can be expressed as $x\omega + y\rho$. 

In addition, We extend our model to a square lattice of size $N = L \times L$ with periodic boundaries and conduct simulations based on the Monte Carlo method. Unlike the well-mixed population, this network exhibits local interactions where individuals only interact with their neighboring nodes. In both the well-mixed population and the square lattice, there are four types of players: weak cooperators ($WC$), weak defectors ($WD$), strong cooperators ($SC$), and strong defectors ($SD$).

\subsection{Evolutionary dynamics in well-mixed population}

In a well-mixed population, each player has the same opportunity to interact with others. Therefore, according to the payoff matrices, the expected payoff of $WC$, $WD$, $SC$, and $SD$ are respectively given by 

\begin{equation}
\begin{split}
\pi_{WC} & = (x \omega  + y \rho)  (b_w - \frac{c}{2}) \\
&+[(1-x)\omega + (1-y) \rho](b_w -c), \\
\pi_{WD} & = (x\omega + y \rho) b_w, \\
\pi_{SC} & = (y \rho + x \omega) (b_s - \frac{c}{2}) \\
 &+[(1-x)\omega + (1-y) \rho](b_s - c), \\
\pi_{SD} &= (x\omega + y \rho)  b_s .	\\
\end{split}
\end{equation}

\begin{figure*}
	\centering
	\includegraphics[scale=9]{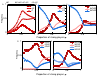}
	\caption{ {\bf Populations of weak and strong cooperators support each other.} A, Cooperation of total population ($TC$) peaks at intermediate densities of weak and strong players. However, cooperation among strong players ($SC$) remains flat even as their proportion increases over intermediate level, while cooperation among weak players ($WC$) collapses together with their proportion. B, Pairwise interactions between cooperators ($CC$) also peaks at the intermediate densities of weak and strong players, but decline as proportion of strong players increases. On the other hand, pairwise interactions between defectors ($DD$) and cooperator-defectors ($CD$) rebound, implying that at higher density, strong players compete more strongly between themselves. C and D, Intraspecific pairwise interactions in both types of players follow the same pattern after proportion of strong players increases beyond intermediate levels, i.e. pairwise cooperation interactions ($WCWC$ and $SCSC$) fall from the most common and are overtaken by interactions between cooperators and defectors ($WCWD$ and $SCSD$). E, Interspecific interactions between cooperators ($WCSC$), however, remain the most abundant even as the proportion of strong players increases, demonstrating a symbiotic relationship. All results are shown for $L=200$, $\alpha$=2, and $b_w$=0.85 with $2 \times 10^5$ time steps.}
	\label{Fig3}
\end{figure*}

We utilize the pairwise comparison with a social learning rule to model the strategy updating stage, where social learning leads to preferential copying of successful strategies in the spread of strategy \cite{szabo2007evolutionary, cardillo2020critical, sigmund2010social}. 
During this stage, a focal player is randomly selected and evaluates its strategy by comparing its payoff with that of another randomly chosen individual in the population. Then, the focal player decides whether to imitate the strategy of the model player, with the probability determined by the Fermi function:

\begin{equation}
\label{Fermi}
W_{\mathcal{X} \leftarrow \mathcal{Y}}(\pi_{\mathcal{X}}, \pi_{\mathcal{Y}}) = \frac1{1+e^{-(\pi_\mathcal{Y}-\pi_\mathcal{X})/K}},
\end{equation}
where $\mathcal{X}$ and $\mathcal{Y}$ represent two different strategies. While the parameter $1/K$ represents selection intensity, adding some noise to the equation in order to allow for a possibility to act irrationally and imitate a worse performing neighbor\cite{szabo2007evolutionary, adami2016evolutionary, szabo2005phase}. Since it has been well studied, we fix $K=0.1$ (a strong selection scenario) throughout this paper. 
In the heterogeneous population, a focal player $WC$ selects $WD$ (or $SD$) as a reference model player for comparison with a probability of $x (1-x)\omega^2 $ (or $ x\omega (1-y)\rho$). Then, the probability that $WC$ decreases by one is
\begin{equation}
\begin{aligned}
Q_1^{-} &=  x (1-x)\omega^2 W_{WC \leftarrow WD} + x \omega (1-y)\rho W_{WC \leftarrow SD}\\
&=  x (1-x)\omega^2 \frac{1}{1+e^{-(\pi_{WD}-\pi_{WC})/K}} \\
& \quad  + x\omega (1-y)\rho \frac{1}{1+e^{-(\pi_{SD}-\pi_{WC})/K}}.
\end{aligned}
\end{equation}
Note that the player's type remains constant during the strategy imitation stage. Therefore, if a weak cooperator imitates the strategy of strong defectors, the weak cooperator becomes a defector, resulting in an increase of $WD$ by one. Conversely, if a weak cooperator imitates the strategy of strong cooperators, the frequency of $WC$ remains constant.
Similarly, if a focal player $WD$ selects a $WC$, $SC$ or $SD$ as reference model players for comparison, the probability that $WC$ increases by one is
\begin{equation}
\begin{aligned}
Q_1^{+} &= x (1-x)\omega^2 W_{WD \leftarrow WC} + y\rho  (1-x)\omega W_{WD \leftarrow SC}\\
&= x (1-x)\omega^2 \frac{1}{1+e^{-(\pi_{WC}-\pi_{WD})/K}} \\
& \quad + y\rho  (1-x)\omega \frac{1}{1+e^{-(\pi_{SC}-\pi_{WD})/K}}.
\end{aligned}
\end{equation}
Subsequently, the dynamics of weak cooperation can be represented by 
\begin{equation}
\begin{aligned}
    \dot{x} & = Q_1^{+} - Q_1^{-}.
\end{aligned}
\end{equation}
Similarly, we can get the dynamics of strong cooperation:
\begin{equation}
\begin{aligned}
    \dot{y} & = y(1-y)\rho^2 \frac{1}{1+e^{-(\pi_{SC}-\pi_{SD})/K}} \\
    & + x\omega (1-y)\rho \frac{1}{1+e^{-(\pi_{WC}-\pi_{SD})/K}}  \\
    & - y(1-y)\rho^2 \frac{1}{1+e^{-(\pi_{SD}-\pi_{SC})/K}} \\
    & - y\rho (1-x) \omega \frac{1}{1+e^{-(\pi_{WD}-\pi_{SC})/K}}.
\end{aligned}
\end{equation}

The feasible solution is determined through numerical iteration of the above system of equations, initialized with $(x,y) \in (0,1)^2$.

\subsection{Agent-based simulation in square lattice}

We have implemented a Monte Carlo simulation to calculate the cooperation rate in a steady state.
Denote $\mathcal{S}_i=(1,0)^T$ or $\mathcal{S}_i=(0,1)^T$ as the cooperation or defection strategy adopted by player $i$. In each time step, randomly selected players play a game with their four nearest neighbors (von Neumann neighborhood) and accumulate a payoff $P_i$. 

\begin{equation} 
\label{equation 1}
P_{i}=\left\{\begin{array}{cc}
\sum_{j \in \Omega_{i}} \mathcal{S}_{i}^{T} M_1 \mathcal{S}_{j}   & \text { if $i$ is a weak player}  \\
\sum_{j \in \Omega_{i}} \mathcal{S}_{i}^{T} M_2 \mathcal{S}_{j}   & \text { if $i$ is a strong player}
\end{array},\right.
\end{equation}
where $\Omega_{i}$ represents the neighbor sets of player $i$. Next, a randomly selected player $j$ in $i$'s neighbor set calculates the payoff similarly. At last, player $i$ imitates the strategy of $j$ with a probability given by the Fermi function. Without the specific declaration, we initialize simulations on a lattice of size $L = 100$, with each player having an equal probability to start off by as a cooperator or defector. The location of weak and strong players on a lattice is determined randomly. In each simulation run, we observe the cooperation frequency as a function of parameters $b_w$ and $\alpha$, and the ratio of weak-to-strong players in the total population over the course of  $10^5$ time steps (without the specific declaration), of which the last 5000 steps are considered as a steady state. In each time step, players are selected once on average to play a game and update their actions.

\section{Results}

\subsection{Well-mixed population}

Cooperation levels in well-mixed populations are significantly influenced by the participation of strong players. In the context of the $PDG$, the composition of the population has minimal impact on global cooperation when both weak and strong players participate (indicated by the red square in Fig.\ref{wellmix}A). However, the incorporation of the $SDG$, particularly with the involvement of strong players, results in the highest levels of cooperation, both in scenarios involving $PDG-SDG$ and $SDG-SDG$ interactions (Fig.~\ref{wellmix}A). Meanwhile, we observe that increasing the value of $b_w$, which weakens the dilemma strength \cite{wang2015universal}, promotes cooperation. This is evident in the evolution dynamics of weak and strong cooperation, as illustrated in Fig.~\ref{wellmix}B and C. The results showcase that when the strong players constitute a small fraction ($\rho=0.2$), weak cooperation is vulnerable to exploitation by weak defectors, leading to a lower level of global cooperation. However, strong cooperators dominate the strong sub-population and provide support for the evolution of weak cooperation. As $\rho$ increases (see Fig.~\ref{wellmix}C), although strong defection emerges, strong cooperation continues positively influencing the evolution of weak cooperators. Interestingly, when the proportion of strong players is small, strong cooperators occupy a higher density among the strong players compared to scenarios with a larger proportion of strong players.

\subsection{Square lattice}

\begin{figure}
	\centering
	\includegraphics[scale=5]{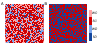}
	\caption{ {\bf Weak players protect cooperation clusters from invasion.} A, in populations of weak and strong players, strong defectors are constrained in small defection clusters, which are in turn held back by strong cooperators. The value of $\rho$ is set as 0.55. B, in a pure population of strong players, cooperators and defectors coexist in numerous clusters. The value of $\rho$ is set as 1. Other parameters are fixed as $\alpha$=2, and $b_w$=0.85.}
	\label{fig4}
\end{figure}

\begin{figure*}
	\centering
	\includegraphics[scale=9]{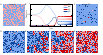}
	\caption{ {\bf Ecological drift between the weak and strong sub-populations fosters higher cooperation.} A, in a pure population consisting of weak players, both cooperators and defectors are randomly assigned within the network initially. C, as time progresses, cooperation experiences a sharp decline. However, the emergence of strong players due to ecological drift alters the dynamics. D, E, and F, cooperation gradually spreads through the formation of clusters, leading to the coexistence of four player types. G, ultimately, four types of players coexist in the population. B, time series clearly depict the evolutionary process of four types of players. The parameters used are $b_w=0.85$, $\alpha=2$, and a mutation rate of 0.01.}
	\label{mufig1}
\end{figure*}

In a square lattice, cooperation emerges as $b_w$ sufficiently increases when the weak and strong players participate in a symmetric game (see Fig.~\ref{Fig1}A). Interestingly, in mixed populations, cooperation levels surpass those in pure populations, especially in scenarios involving intermediate inequality degrees in both $PDG-SDG$ and $SDG-SDG$ scenarios (see Fig.~\ref{Fig1}B and C). This is not surprising when comparing populations of pure weak players, as the presented values of $b$ place the social dilemma in $PDG$ in the environment where only defection can survive \cite{szabo2005phase}. However, it seems counter-intuitive when compared to populations of pure strong players. There, cooperation survives and even thrives at low enough values of parameter $b$. Even so, cooperation further increases with an influx of weak players and it is at highest when the proportion of weak and strong players is at intermediate levels. This phenomenon is particularly significant at the intermediate value of inequality degree $\alpha$. However, as $\alpha$ increases further, cooperation in mixed populations ceases to surpass that in pure populations of strong players (see Fig.~\ref{Fig1}D, E, and F). It's noteworthy that this optimal effect on global cooperation diminishes first in the $SDG-SDG$ scenario, followed by the $PDG-SDG$ scenario, with the increase of the asymmetric degree. Global cooperation monotonously increases with the proportion of strong players when the asymmetric degree is sufficiently large. Time series analysis reveals pretty straightforward evolutionary dynamics (see Fig.~\ref{FigA1} in Appendix), with strong and weak cooperation as the most dominant strategies in high cooperation scenarios (see Fig.~\ref{FigA1}C), followed by weak and strong defection. Specifically, Fig.~\ref{FigA1}C-E demonstrate that the temporal evolution of weak cooperators and strong cooperators initially follows a downward trend, followed by an upward trajectory, indicative of the presence of network reciprocity \cite{perc2008restricted, szolnoki2009promoting}.

We, therefore, turn to the analysis of pairwise interactions between players (Fig.~\ref{Fig3}). Total cooperation peaks when the proportion of strong and weak players is at intermediate level (Fig.~\ref{Fig3}A), and decreases when the proportion of strong players is increased beyond that. However, the cooperativeness of strong players remains mostly flat after the peak at intermediate density, while the cooperativeness of weak players diminishes together with their density, implying that they are overtaken by defectors. At $\rho=0$, the population is entirely composed of weak players, resulting in predominant weak defection. Conversely, in a population consisting solely of strong players, we observe a mix of cooperation and defection. Apart from these two extremes, all four states (weak cooperator, strong cooperator, weak defector, and strong defector) coexist. Meanwhile, the frequency of pairwise interactions between defectors increases with the proportion of strong players going beyond intermediate level (Fig.~\ref{Fig3}B), while pairwise interactions between cooperators and defectors become the most abundant at high densities of strong players (Fig.~\ref{Fig3}B-D). Looking at the interspecific pairwise interactions further untangles the propensity for cooperation at intermediate densities of players, as more than half of all interactions are between strong and weak cooperators (Fig.~\ref{Fig3}E, F). Next in the abundance are interactions between weak defectors and strong cooperators, weak and strong defectors, and finally weak cooperators and strong defectors.

In order to explain the optimal effect triggered by mixed populations with inequality from an individual level, we scrutinize the snapshots in steady state given by Fig.~\ref{fig4}. Contrary to previous research that identified cyclical dominance among coexisting states \cite{szolnoki2010dynamically}, the evolutionary dynamics of $WC$, $WD$, $SC$, and $SD$ in our study do not form a closed loop due to the constancy of individual tags. As a result, when the population approaches an asymptotically stable state, strategy imitation predominantly occurs among minority players situated at the boundary of strategy clusters. Steady-state snapshots limn the clearer picture: mixed populations enable weak players to insulate strong defectors from cooperative clusters, hence protecting strong cooperators from the standard dynamics of $SDG$ in pure strong populations (Fig.~\ref{fig4}). This isolation strip decreases after cooperation peaks, which sufficiently explains the observed phenomenon. So far, we have been focused on the environments with constrained population dynamics, i.e. the densities of weak and strong players have been kept fixed. However, to test how the evolution of cooperation changes if there is a transition between weak and strong players, we modified our model to include ecological processes. In contrast to the strategy mutation during the updating stage \cite{helbing2010defector}, we allow for population tag to change via ecological drift, i.e., a small, fixed probability at the beginning of each round for weak players to change to strong players and vice versa, and initialize the model with a population of pure weak players (see Fig.~\ref{mufig1}). Initially, the game unfolds with significantly reduced cooperation (Fig.\ref{mufig1}A-C). However, when strong cooperators emerge through mutation, cooperation resists the invasion of defection and spreads in the network (Fig.\ref{mufig1}D-F). Simultaneously, strong defection emerges through mutation and the invasion of cooperators. In the end, cooperation accounts for a larger proportion than defection (Fig.\ref{mufig1}G), which is consistent with the results obtained in the scenario $\rho=0.5$ without mutations. Furthermore, we conduct sensitivity analysis by considering two scenarios: one with a pure strong population initially and the other with a mixed population of weak and strong initially. Remarkably, both scenarios reach identical results (see Fig.~\ref{figA2} in Appendix).

\section{Conclusion}

We study the joint effect of population composition and unequal benefits on the evolution of human cooperation, utilizing both structured and unstructured populations. The population consists of strong and weak players, where the former consistently gains a larger benefit from cooperation compared to the latter, indicating an asymmetric degree. When cooperators and defectors are assigned to strong and weak players, four types of players emerge, i.e., strong cooperators, strong defectors, weak cooperators, and weak defectors. In consideration of the benefits for weak players and the asymmetric degree, three types of mixed games are identified: $PDG-PDG$, $PDG-SDG$, and $SDG-SDG$ mixed scenarios. To calculate cooperation in steady states, we employ mean-field theory for well-mixed populations and Monte Carlo simulations for structured populations.

In well-mixed populations, an increased proportion of strong players is more likely to result in higher levels of cooperation. Maximum cooperation is achieved in a pure strong population when $SDG$ is incorporated. However, when networks are considered, the aforementioned findings may differ. A structured population with a moderate proportion of strong players triggers the highest level of global cooperation, particularly when the asymmetric degree is at a moderate level. This phenomenon can be attributed to the distribution of players within the network, where weak players insulate strong defectors from cooperative clusters, shielding them from the standard dynamics of the $SDG$ found in pure populations. Nevertheless, if the asymmetric degree becomes sufficiently large, global cooperation increases monotonously with the proportion of strong players, as they participate in the game with lower dilemma strength.

Our research contributes significantly to the comprehension of asymmetric evolutionary games, particularly focusing on the interplay between inequality and population composition. However, we recognize some constraints in our approach. The use of two-player games, commonly adopted for examining the evolution of cooperation, tends to oversimplify the complexities inherent in real-world interactions. In comparison, $N$-player public goods games, which are another archetypal representation of social dilemmas, are more apt for modeling intricate interaction scenarios \cite{santos2011risk, chen10136810}. Consequently, broadening our study to encompass $N$-player collective games could yield valuable perspectives on fostering optimal cooperation within heterogeneous populations. Additionally, considering time scales or time delays, which capture the fast and slow effects between strategy updating and identity tag updating, can lead to more realistic analyses \cite{rong2013coevolution}. On the other hand, the precision of experimental investigations in capturing human decision-making processes offers a valuable opportunity \cite{nishi2015inequality, li2018punishment}. Organizing human experiments in this domain would not only provide validation but also enhance our comprehension of asymmetric interactions, bringing a more realistic perspective to our studies. We believe that these findings will continue to be useful for explaining and manipulating the evolution of cooperation in scenarios with asymmetry in real life.

\bigskip

\section*{Acknowledgement}
This research was supported by the National Science Fund for Distinguished Young Scholars (No. 62025602), the National Science Fund for Excellent Young Scholars (No. 62222606), the National Natural Science Foundation of China (Nos. 11931015, U1803263, 81961138010 and 62076238), Fok Ying-Tong Education Foundation, China (No. 171105), Technological Innovation Team of Shaanxi Province (No. 2020TD-013), Fundamental Research Funds for the Central Universities (No. D5000211001), the Tencent Foundation and XPLORER PRIZE, JSPS Postdoctoral Fellowship Program for Foreign Researchers (grant no. P21374), and China Postdoctoral Science Foundation (No. 2023M741852).

\section*{Authors’ contributions}
H.G.: Conceptualization, formal analysis, investigation, methodology, visualization, writing-original draft, writing-review and editing; C.S.: formal analysis, investigation, writing-original draft, writing-review and editing; R.Z.: investigation, resources, visualization; P. T: supervision, writing-review and editing; Y.S: supervision, writing-review and editing; Z.W.: project administration, supervision, writing-original draft, writing-review and editing; J.X.: conceptualization, supervision, writing-original draft, writing-review and editing.

\section*{Author Declarations}
\subsection*{Conflict of interest}
The authors have no conflicts to disclose.

\section*{Data AVAILABILITY}
The data that support the findings of this study are available from the corresponding author upon reasonable request.

\appendix

\section*{Appendix}

\begin{figure}[h]
	\centering
	\includegraphics[scale=10]{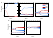}
	\caption{ {\bf Time series of weak cooperators, weak defectors, strong cooperators, and strong defectors.} A-E depict the evolutionary games with varying proportions of strong players: $0$, $25\%$, $50\%$, $75\%$, and $100\%$, respectively. In a population of pure weak players (A), cooperation cannot survive, while in a population of pure strong players (E), cooperation and defection coexist at similar levels. In a population with a low proportion of strong players (B), defection is dominant but cooperation manages to survive. However, at intermediate proportion (C) cooperation is dominant both in weak and strong populations, while at high proportion (D) cooperation is more common than defection in both weak and strong populations. Variables are set as $\alpha$=2 and $b_w$=0.85.}
	\label{FigA1}
\end{figure}

\begin{figure}[h]
	\centering
	\includegraphics[scale=5]{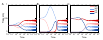}
	\caption{ {\bf Evolution steady state in ecological drift is insensitive to the initial frequency of each player.} 
    A, each player is assigned to be a cooperator with a probability of 0.5. Furthermore, each player is randomly assigned to either the weak tag or the strong tag with an equal probability of 0.5. B and C, no strong and weak players are included in the network initially. The parameters used in these simulations include $b_w = 0.85$, $\alpha = 2$, and a mutation rate of 0.01.
	}
	\label{figA2}
\end{figure}


\section*{REFERENCES}


%

\end{document}